# Band structure of tetragonal and orthorhombic fluorine-arsenide SrFeAsF as a parent phases for a new group of oxygen-free FeAs superconductors

**I.R. Shein * and A.L. Ivanovskii**

*Institute of Solid State Chemistry of the Ural Branch of the Russian Academy of Sciences,
620041, GSP-145, Ekaterinburg, Russia*

**Abstract**

The full-potential linearized augmented plane wave method with the generalized gradient approximation for the exchange and correlation potential (LAPW-GGA) is used to understand the electronic band structure of fluorine-arsenide SrFeAsF as a possible parent material for a new group of oxygen-free FeAs superconductors. The electronic bands, density of states, Fermi surface and atomic charges have been evaluated and discussed for high-temperature tetragonal and low-temperature orthorhombic SrFeAsF phases.

*Corresponding author*: e-mail shein@ihim.uran.ru,
Phone: +7 343 362 31 15, Fax: +7 343 374 44 95

The surprising discovery [1-7] of new copper-free high-temperature FeAs superconductors (SCs) with the transition temperatures to $T_C \sim 56K$ has stimulated much activity in search for new related superconducting materials, and element substitution seems to be an efficient way to produce new SCs of this family. However until now most of the research groups are focusing on the most numerous group of these SCs based on oxygen – containing layered quaternary oxyarsenides *Ln*FeAsO, (where *Ln* = La, Ce … .Gd, Tb, Dy), which adopt a tetragonal (space group P4/*nmm*) quasi-two-dimensional crystal structure.

Quite recently, the new isostructural oxygen-free quaternary phases – *A*FeAsF (where *Ln* = Sr, Ca and Eu) have been synthesized [8-12] and some of their properties have been determined. In particular, for these compounds the strong temperature-dependent anomalies in the specific heat, electrical resistance and magnetic susceptibility (which were interpreted as spin density wave (SDW) anomalies) were observed [8-12]. Similar SDW anomalies are known to be an important prerequisite for high-$T_C$ superconductivity in oxygen-containing FeAs materials.

Moreover, critical transitions with $T_C \sim 4K$ in the Co- substituted SrFeAsF ($SrFe_{0.875}Co_{0.125}AsF$) [9], $T_C \sim 22K$ in the Co- substituted CaFeAsF ($CaFe_{0.9}Co_{0.1}AsF$) [12], and $T_C \sim 31K$ in the La- substituted SrFeAsF ($Sr_{0.8}La_{0.2}FeAsF$) were reported [11]. These results clearly show that *A*FeAsF phases can serve as prototype materials for a new group of oxygen -free FeAs SCs.

The available data [8] reveal that at room temperature the synthesized SrFeAsF samples adopt a tetragonal symmetry (space group P4/*nmm*) while XRD patterns below 185 K were indexed with an orthorhombic unit cell (space group C*mme*).

In this Communication, using the first principles FLAPW method within the generalized gradient approximation (GGA) for the exchange-correlation potential we explore for the first time the electronic properties for both tetragonal (T) and orthorhombic (O) phases of fluorine-arsenide SrFeAsF.

Our calculations were performed by means of the full potential all-electron DFT method with the mixed basis APW+lo (LAPW) method implemented in the WIEN2k code [13]. The generalized gradient approximation (GGA) to the exchange-correlation potential [14] was used. The calculations were performed with full-lattice optimizations including the so-called internal parameters $z_{Sr}$ and $z_{As}$. The self-consistent calculations were considered to be converged when the difference in the total energy of the crystal did not exceed 0.1 mRy and the difference in the total electronic charge did not exceed 0.001 *e* as calculated at consecutive steps.

The hybridization effects were analyzed using the densities of states (DOSs), which were obtained by a modified tetrahedron method [15]. The Bader [16] analysis was employed to discuss the ionic bonding..

As the first step, the total energy ($E_{tot}$) *versus* cell volume calculations were carried out to determine the equilibrium structural parameters for T and O phases of SrFeAsF; the calculated values are presented in Table 1 and are in reasonable agreement with the available experiments [8-11].

The calculated band structures and DOSs pictures demonstrate close similarities for T and O phases. Therefore let us discuss the main peculiarities of the electronic structure of SrFeAsF using the T-phase as an example, see Figs. 1-3.

Figure 1 shows the band structure of T-SrFeAsF as calculated along the high-symmetry *k* lines. As can be seen, the quasi-core As 4*s* bands are located from -12.4 eV up to -10.6 eV below the Fermi level ($E_F$) and are separated from the valence band (VB) by a gap. The VB extends from -6.8 eV up to the Fermi level $E_F$ = 0 eV and includes three main subbands *A-C*, Fig. 2. The first subband *A* is formed mainly by F 2*p* states with some admixture of Sr *sp* states, whereas the subband *B* is formed predominantly by hybridized As 4*p* and Fe 3*d* states. These bands are responsible for Fe-As and As-As covalent bonds owing to hybridization of Fe 3d- As 4p and As 4p- As 4p states, respectively. The near-Fermi subband *C* contains the main contributions from the Fe 3*d* states which form Fe-Fe bonds.

The most interesting feature of the band structure of SrFeAsF is the behavior of quasi-flat electronic bands along Γ-Z and A-M directions in the Brillouin zone (BZ). The corresponding Fermi surface (FS) in the first BZ is shown in Fig. 3. Owing to the quasi-two-dimensional electronic structure, the Fermi surface is composed of cylindrical-like sheets which are parallel to the $k_z$ direction. Three of them are of a hole-like character and are centered along the Γ - Z high symmetry lines, whereas the other two sheets are of the electronic-like type and are aligned along the A - M direction. Note that similar FS topology was established also for the oxygen-containing FeAs SCs, see [17].

To define the comparative stability of the O and T phases, we have calculated their total energy difference $\Delta E_{tot}$. In this way, we have obtained the $\Delta E_{tot}$ value of about 0.01 eV/cell, which shows, in agreement with the experiment [8], that the most stable phase is the orthorhombic phase, whereas the tetragonal SrFeAsF is a high-temperature polymorph.

As electrons near the Fermi surface are involved in the formation of the superconducting state, it is important to figure out their nature. The total and orbital decomposed partial DOSs at the Fermi level, $N(E_F)$, are shown in Table 2. It is seen that for both O and T SrFeAsF phases the main contribution to $N(E_F)$ is from the Fe 3*d* states, and $N(E_F)$ decreases slightly as going from the T to the O phase.

For the oxygen-containing FeAs superconducting materials it is established that magnetic fluctuations in the iron layers close to the quantum critical point play a fundamental role in the superconducting pairing mechanism [1-7,16]. Therefore we used our total energy calculations to examine the magnetic instability of low-temperature O-SrFeAsF for which the non-magnetic (NM) and magnetic (in assumption of ferromagnetic (FM) spin ordering) calculations were performed. Our results showed that the energy difference between NM and FM states is very small (~ 0.1 mRy/cell), indicating that SrFeAsF lies at the border of magnetic instability. The values of $N(E_F)$ are 1.527 states/eV·atom$^{-1}$ and 1.397 states/eV·atom$^{-1}$ for NM and FM states, *i.e.* they decrease as going from the NM to the magnetic state of O-SrFeAsF. The calculated local magnetic moments of Fe atoms for the ferromagnetic state of O-SrFeAsF are about 0.13 $\mu_B$.

Finally let us discuss shortly the peculiarities of the chemical bonding for SrFeAsF. The idealized picture assumes a purely ionic bonding between the atoms (inside [SrF] and [FeAs] layers) as well as between the adjacent layers [SrF]/[FeAs] according to ionic formula $[Sr^{2+}F^{1-}]^{1+}[Fe^{2+}As^{3-}]^{1-}$. To estimate the amount of electrons redistributed between the adjacent [SrF]/[FeAs] layers and between the atoms inside each layer, we carried out Bader analysis. The total charge of an atom (the so-called Bader charge, $Q^B$), as well as the corresponding charges as obtained from the purely ionic model ($Q^i$) and their differences ($\Delta Q = Q^B - Q^i$) are presented in Table 3. These results show that the inter-layer charge transfer is much smaller than it is predicted in the idealized ionic model. Namely, the transfer $\Delta Q([SrF] \rightarrow [FeAs])$ for SrFeAsF is about 0.31 e, and this value is smaller than for LaFeAsO (0.39 e [16]). In addition, the orbital mixing picture is evident from the valence DOSs shapes, see above.

As a result, the bonding in SrFeAsF can be classified as a mixture of metallic, ionic and covalent contributions, namely: (1). inside [SrF] layers, the ionic Sr-F bonds take place with a small admixture of hybridization of valence states of Sr-F; (2). inside [FeAs] layers, mixed metallic-ionic-covalent Fe-As bonds appear (owing to hybridization of valence states of Fe-Fe and Fe-As atoms and Fe $\rightarrow$ As charge transfer); in addition, inside [FeAs] layers covalent bonds As-As take place (owing to As $4p$ - As $4p$ hybridization); and (3) between the adjacent [SrF] and [FeAs] layers, ionic bonds emerge owing to [SrF] $\rightarrow$ [FeAs] charge transfer, and these bonds are responsible for the cohesive properties of the SrFeAsF crystal. Thus, SrFeAsF may be described as a *quasi-two-dimensional ionic metal*.

In conclusion, we used the first-principle FLAPW-GGA approach to investigate the structural, electronic properties and stability of tetragonal and orthorhombic polymorphs of fluorine-arsenide SrFeAsF as a possible parent material for a new group of oxygen-free FeAs SCs. We found that the O-phase is more favorable than SrFeAsF with a tetragonal structure.

On the other hand our calculations show that their band structures and DOSs pictures are very similar. The most interesting feature of the band structure of SrFeAsF is the quasi-flat Fe 3*d*-like electronic bands which form the characteristic Fermi surface consisting of a set of cylindrical-like sheets parallel to the $k_z$ direction. In addition, we found that SrFeAsF: (1) is at the border of magnetic instability (2). adopts a complex metallic-ionic-covalent bonding picture, where the ionic bonds between the adjacent [SrF] and [FeAs] layers are responsible for the cohesive properties of this crystal.

The mentioned peculiarities of electronic properties of SrFeAsF reveal close similarity for the oxygen-containing FeAs SCs [1-7,16], which may be considered as the theoretical background determining the possibility of superconductivity in fluorine-arsenide compounds. Naturally, further in-depth studies are necessary to understand the scenarios of superconducting coupling mechanisms for these new oxygen-free materials.

**Table 1.** The optimized lattice parameters ($a$ and $c$, in Å), internal coordinates ($z_{Sr(Ca)}$ and $z_{As}$), some inter-atomic distances ($d$(Fe-As), in Å) and angles (As-Fe-As, in deg.) and cell volumes ($V_o$, in Å$^3$) for tetragonal (space group P4/*nmm*) and orthorhombic (space group C*mme*) SrFeAsF phases as compared with available experiments.

| Phase /parameter [1] | tetragonal (s. g. P4/*nmm*) | orthorhombic (s. g. C*mme*) |
|---|---|---|
| $a$ | 4.0055 (3.9930 [8]; 3.999 [9]; 4.004 [10]; 4.011 [11]) | 5.6623 (5.6155 [8]) |
| $b$ | = $a$ | 5.6640 (5.6602 [8]) |
| $c$ | 8.8049 (8.9546 [8]; 8.973 [9]; 8.971 [10]; 8.965 [11]) | 8.8575 (8.9173 [8]) |
| $c/a$ | 2.1982 (2.2426 [8]; 2.2438 [9]; 2.2405 [10]; 2.2351 [111]) | 1.5675 (1.5880 [8]) |
| $z_{As}$ | 0.6397 (0.6527 [8]) | 0.6384 (0.6494 [8]) |
| $z_{Sr,Ca}$ | 0.1665 (0.1598 [8]) | 0.1663 (0.1635 [8]) |
| $d$(Fe-As) | 2.350 (2.420 [8]) | 2.348 (2.397 [8]) |
| (As-Fe-As) | 105.89 (108.6 [8]) | 105.8-117.1 (107.7-112.5 [8]) |
| $V_o$ | 141.27 (142.77 [8]) | 284.07 (283.43 [8]) |

[1] the available experimental data are given in parentheses.

**Table 2.** Total $N^{tot}(E_F)$ and partial $N^l(E_F)$ densities of states at the Fermi level (in states/eV·atom$^{-1}$) for tetragonal (space group P4/*nmm*) and orthorhombic (space group C*mme*) SrFeAsF phases.

| Phase /parameter | tetragonal (s. g. P4/*nmm*) | orthorhombic (s. g. C*mme*) [1] | |
|---|---|---|---|
| | | NM | FM |
| $N^{Fed}(E_F)$ | 1.188 | 1.173 | 1.075 |
| $N^{As}(E_F)$ | 0.039 | 0.083 | 0.071 |
| $N^{tot}(E_F)$ | 1.540 | 1.527 | 1.397 |

[1] for non-magnetic (NM) and ferromagnetic (FM) states

**Table 3.** Effective atomic charges and charges of [FeAs] and [SrF] layers (in e) for T-SrFeAsF as obtained from a purely ionic model ($Q^i$), Bader analysis ($Q^B$) and their differences ($\Delta Q = Q^B - Q^i$) in comparison with LaFeAsO [18].

| phase | Q | Sr(La) | Fe | As | F(O) |
|---|---|---|---|---|---|
| SrFeAsF | $Q^i$ | +2 | +2 | -3 | -1 |
| (s. g. | $Q^B$ | 8.465 | 7.741 | 5.950 | 7.843 |
| P4/nmm) | $\Delta Q$ | 0.465 | 1.741 | -2.050 | -0.157 |
| LaFeAsO | $Q^i$ | +3 | +2 | -3 | -2 |
| (s. g. | $Q^B$ | 9.115 | 7.722 | 5.887 | 7.276 |
| P4/nmm) | $\Delta Q$ | 1.115 | 1.772 | -2.113 | -0.724 |
| phase | Q | [Sr(La)F(O)] | | [FeAs] | |
| SrFeAsF | $Q^i$ | +1 | | -1 | |
| (s. g. | $Q^B$ | 16.309 | | 15.585 | |
| P4/nmm) | $\Delta Q$ | 0.309 | | -0.309 | |
| LaFeAsO | $Q^i$ | +1 | | -1 | |
| (s. g. | $Q^B$ | 16.391 | | 14.998 | |
| P4/nmm) | $\Delta Q$ | 0.391 | | -0.391 | |

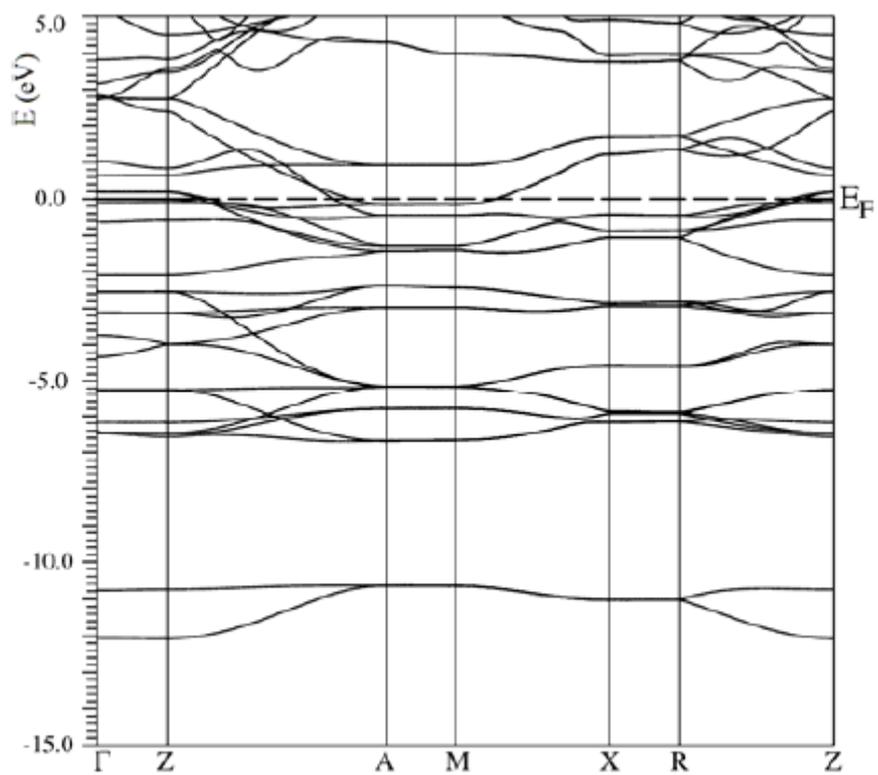

**Figure 1.** Electronic band structure of T-SrFeAsF.

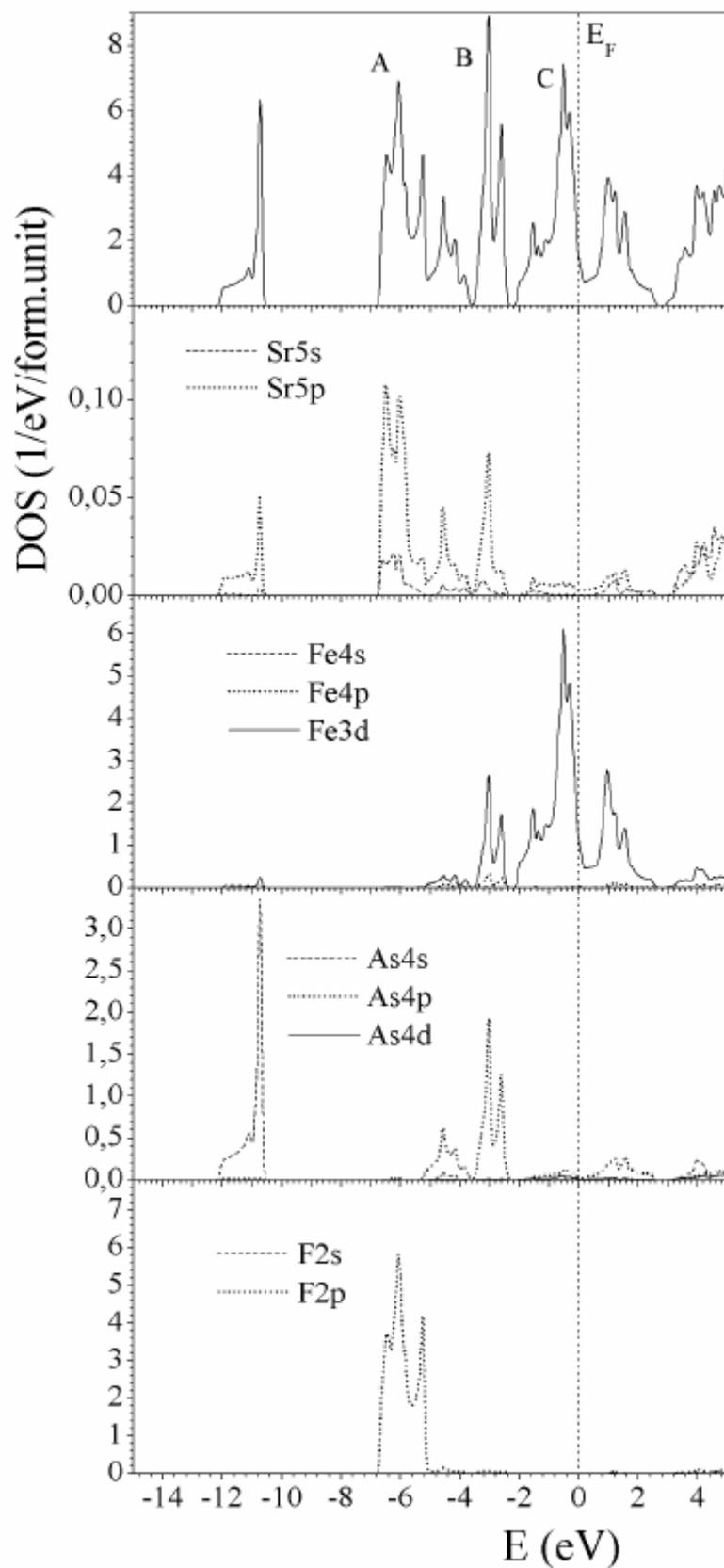

**Figure 2.** The total and partial DOSs for T-SrFeAsF.

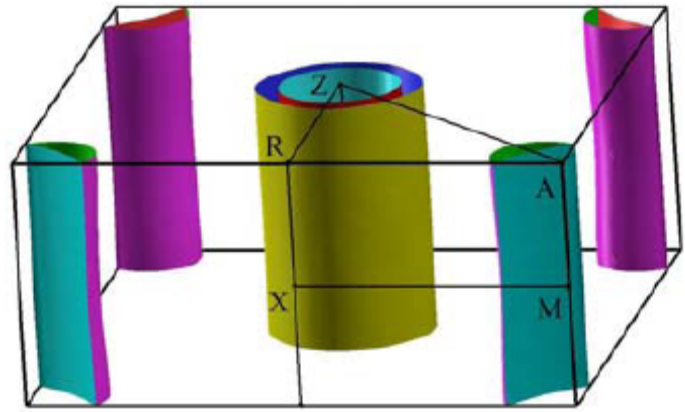

**Figure 3.** The Fermi surface for T-SrFeAsF.